\documentclass[prodmode,license]{acmsmall-ec15}

\doi{2764468.2764488}

\usepackage[ruled]{algorithm2e}
\renewcommand{\algorithmcfname}{ALGORITHM}
\SetAlFnt{\small}
\SetAlCapFnt{\small}
\SetAlCapNameFnt{\small}
\SetAlCapHSkip{0pt}
\IncMargin{-\parindent}

\usepackage[numbers]{natbib}

\PassOptionsToPackage{hyphens}{url}\usepackage{url}
\usepackage{comment}
\usepackage{graphicx}
\usepackage{grffile}
\usepackage{verbatim}
\usepackage[caption=false]{subfig}

\newcommand{\sref}[1]{Section~\ref{#1}}
\newcommand{\eref}[1]{Equation~\ref{#1}}
\newcommand{\fref}[1]{Figure~\ref{#1}}

\newcommand{\DfDx}[2]{\textstyle \frac{\mathrm{d}\!#1}{\mathrm{d}\!#2}}

\newcommand{\tz}{t_0}
\newcommand{\tb}{t_-}
\newcommand{\ts}{t_*}

\newcommand{\uit}{\mathrm{u}_{it}}
\newcommand{\ujt}{\mathrm{u}_{jt}}

\newcommand{\vi}{\mathrm{v}_{i}}
\newcommand{\vit}{\mathrm{v}_{it}}
\newcommand{\vits}{\mathrm{v}_{i\ts}}
\newcommand{\vitb}{\mathrm{v}_{i\tb}}
\newcommand{\vjt}{\mathrm{v}_{jt}}

\newcommand{\djt}{\mathrm{d}_{jt}}

\newcommand{\rij}{\mathrm{r}_{ij}}
\newcommand{\rijt}{\mathrm{r}_{ijt}}

\newcommand{\rkjt}{\mathrm{r}_{kjt}}
\newcommand{\rijts}{\mathrm{r}_{ij\ts}}
\newcommand{\rijtb}{\mathrm{r}_{ij\tb}}

\newcommand{\vpit}{\mathrm{v}'_{it}}
\newcommand{\cijt}{\mathrm{c}_{ijt}}

\newcommand{\cjtb}{\mathrm{c}_{j\tb}}

\newcommand{\rhoij}{\rho_{ij}}
\newcommand{\rhoijhat}{\hat{\rho}_{ij}}
\newcommand{\rhoihat}{\hat{\rho}_{i}}

\newcommand{\sigmaij}{\sigma_{ij}}
\newcommand{\lambdaij}{\lambda_{ij}}

\newcommand{\lambdaoverall}{\hat{\lambda}}

\begin{document}

\title{Estimating the Causal Impact of Recommendation Systems from Observational Data}
\markboth{Sharma et al.}{Estimating the causal impact of recommendation systems from observational data}

\author{
AMIT SHARMA
\affil{Cornell University}
JAKE M. HOFMAN
\affil{Microsoft Research}
DUNCAN J. WATTS
\affil{Microsoft Research}
}

\begin{abstract}
Recommendation systems are an increasingly prominent part of the web, accounting for up to a third of all traffic on several of the world's most popular sites.
Nevertheless, little is known about how much activity such systems actually \emph{cause} over and above activity that would have occurred via other means (e.g., search) if recommendations were absent.
Although the ideal way to estimate the causal impact of recommendations is via randomized experiments, such experiments are costly and may inconvenience users. In this paper, therefore, we present a method for estimating causal effects from purely observational data.
Specifically, we show that causal identification through an instrumental variable is possible when a product experiences an instantaneous shock in direct traffic and the products recommended next to it do not.
We then apply our method to browsing logs containing anonymized activity for 2.1 million users on Amazon.com over a 9 month period and analyze over 4,000 unique products that experience such shocks.
We find that although recommendation click-throughs do account for a large fraction of traffic among these products, at least 75\% of this activity would likely occur in the absence of recommendations.
We conclude with a discussion about the assumptions under which the method is appropriate and caveats around extrapolating results to other products, sites, or settings.
\end{abstract}


\keywords{recommender systems; casual inference; natural experiments; log data}

\category{J.4}{Social and Behavioral Sciences}{Economics}


\begin{bottomstuff}
Author's addresses: A. Sharma, Dept. of Computer Science,
220 Gates Hall, Cornell University, Ithaca, NY, 14850, USA ; J. M. Hofman {and} D. J. Watts,
Microsoft Research, 641 Ave of the Americas, 7th Floor, New York, NY 10011, USA.
\end{bottomstuff}

\maketitle

\section{Introduction}

How much activity do recommendation systems cause?
At first glance, answering this question may seem straightforward: given browsing data for a web site, simply count how many pageviews on the site come from clicks on recommendations and compare this to overall traffic. Indeed, exercises of precisely this sort have been conducted~\cite{mulpuru:2006,grau:2009,sharma:2013}, leading to estimates that recommenders generate roughly 10-30\% of site activity and revenue.
But these estimates likely overstate the true causal estimate, possibly by a large amount. To see why, consider users who visit Amazon.com in search of a pair of winter gloves. 
Upon viewing the product page for the gloves, some users might notice a winter hat listed as a recommendation and click on it to continue browsing. According to the naive approach that simply counts clicks, this view would be attributed to the recommender system. But the question we focus on here is whether the recommender {\it caused} these users to view another product---in this case a winter hat---or if they would have done so anyway in a counterfactual world in which the recommender did not exist~\cite{rubin:2005}.
In this example it seems quite likely that users looking for winter gloves would be interested in winter clothing in general.
In the absence of a recommendation, therefore, such a user might well have conducted a separate search for a winter hat and ended up on the same page regardless; thus the recommender could not be said to have caused the visit in the strict counterfactual sense.

This example highlights the problem of correlated demand: if interest in a product and its recommendations are correlated, then simply counting recommendation click-throughs overestimates the number of views caused by recommendations.
Moreover, because such correlations are likely to be common---indeed, systems such as Amazon's ``Customers who bought this also bought''~\cite{linden:2003} rely on them to generate their recommendations---the overestimate is potentially large.
One could, of course, control for correlated demand by running experiments in which recommendations were randomly turned on or off throughout the site to obtain causal estimates.
Past work in this direction confirms the above intuition, indicating substantially lower estimates of the impact of recommendation systems~\cite{dias:2008,belluf:2012,jannach:2009}.
Unfortunately, experiments of this sort are costly to run in terms of time or revenue and may also negatively impact user experience.

An alternative route is therefore to identify natural experiments in observational data that can be used to estimate causal effects~\cite{angrist:2008,jensen:2008,oktay:2010,dunning:2012}.
In this approach one looks for naturally occurring events that simulate random assignment, effectively decoupling variables that might otherwise be correlated.
One such natural experiment involves looking at products that experience large and sudden increases in traffic and counting the number of associated recommendation click-throughs~\cite{carmi:2012,kummer:2013}.
The hope is that such ``exogenous shocks'' are analogous to a controlled experiment in which the experimenter randomly exposes people to product pages and measures resulting recommendation activity.

Unfortunately, natural experiments involving exogenous shocks do not necessarily solve the problem of correlated demand either. Consider, for example, the book ``Tenth of December'' written by George Saunders, who appeared on the Colbert Report in January of 2013 to promote its release.
The product page for this book on Amazon.com lists a number of similar items along side it, including several of Saunders' other popular works such as ``CivilWarLand in Bad Decline'' and ``Pastoralia''.
Many individuals visited ``Tenth of December'' after it was featured on the show, and some of them clicked through on these recommendations.
As with the winter clothing example, however, Saunders' appearance on Colbert might have increased interest in his books in general, hence some of the viewers of the recommended books might have discovered them anyway through some other means (e.g., search), even had they not been exposed to recommendations.
Past work attempts to control for this by conditioning on observable covariates or comparing activity to a set of ``complimentary'' products~\cite{oestreicher:2012,carmi:2012}, but the success of these approaches can be difficult to verify.

The ideal natural experiment, therefore, is one in which we not only see an exogenous shock to demand for a particular ``focal'' product, but where we also know that demand for a corresponding recommended product is constant.
In the language of causal inference, a shock to the focal product can be treated as an \emph{instrumental variable} \cite{dunning:2012, morgan:2007} to identify the causal effect of the recommendation. 
When the demand for the recommended product is known to be constant, any increase in click-throughs from the focal product can be attributed to the recommender, and hence we can estimate its causal effect simply by dividing the observed change in recommendation click-throughs during the shock by the exogenous change in traffic over the same period.

The main contribution of this paper is to formalize and justify the conditions for such an idealized experiment, and to present a method for constructing instrumental variables of this sort from log data.
Specifically, the remainder of the paper proceeds as follows. First, we review related work (\sref{sec:related}) and then describe our data (\sref{sec:data}), comprising 23 million visits by 2.1 million Bing toolbar users to 1.38 million Amazon.com products over a nine month period.
Next, in \sref{sec:methods} we present a formal causal model describing recommendation click-throughs, and use this model to derive a simple estimator for the causal impact of recommendations under certain assumptions that we specify.
Also in \sref{sec:methods} we specify a set of heuristics for finding products that receive shocks while the products recommended next to them do not, and identify over 4,000 experiments that satisfy our criteria.
Next, in \sref{sec:results} we use our method to show that although recommendation click-throughs do account for a large fraction of traffic among these products, at least 75\% of this activity would likely occur in the absence of recommendations---a number that corresponds surprisingly well with estimates from a recent field experiment~\cite{belluf:2012}.
Finally, in \sref{sec:discussion} we discuss some limitations to our method, but also emphasize that although our results are specific to Amazon's recommendation system, the methods we develop are general and can be applied whenever one has access to data that log the number of recommendation-driven visits and number of total visits to individual pages over time.


\section{Related Work}
\label{sec:related}
There is an extensive body of work on recommender systems that seeks to evaluate such systems along various metrics including accuracy, diversity, utility, novelty and serendipity of the recommendations shown to users \cite{herlocker:2004,mcnee:2006,shani:2011}. Among these many possible dimensions of recommender systems, we focus specifically on the role of recommendations in exposing users to items they would not have seen otherwise---a function that is closely related to the notion of serendipity, defined as recommending a ``surprisingly interesting item a user might not have otherwise discovered" \cite{herlocker:2004}---and thus, \textit{causing} an increase in the volume of traffic on a website. Although our somewhat narrow focus on increasing volume clearly overlooks other potentially important functions of recommenders, it greatly simplifies the methodological challenges associated with estimating causal effects, allowing us to make progress. 

Focusing specifically on volume, therefore, previous work on estimating the impact of recommendation systems can be classified into two broad categories: experimental and non-experimental approaches.
In the experimental category, Dias et al.~\citeyearpar{dias:2008} tracked usage of a recommendation system on a Swiss online grocer over a two year period following its introduction in May 2006, finding that both click-throughs and associated revenues increased over the study interval. Because they did not compare either total pageviews or revenue with a control condition (i.e., without recommendations), however, it is impossible to estimate how much of this increase was caused by the recommendation system itself versus some other source of demand. Subsequently, Jannach and Hegelich~\citeyearpar{jannach:2009} randomly assigned 155,000 customers of a mobile game platform to see either personalized or non-personalized recommendations, finding that personalized recommendations generated significantly more clicks and downloads than non-personalized recommendations. Compared with a prior no-recommendation condition, moreover, they estimated that personalized recommendations could have increased sales by as much as 3.6\%.  Finally, Belluf et al.~\citeyearpar{belluf:2012} conducted an experiment on a Latin American shopping website in which 600,000 users were randomly assigned to either receive or not receive recommendations for one month in 2012, finding that recommendations increased pageviews per user by 5-9\%. 

In the non-experimental category, Garfinkel et al.~\citeyearpar{garfinkel:2006} analyzed panel data comprising 156 books on Amazon.com and Barnes and Noble over a 52 day period. By conditioning on observable covariates, including previous day sales rank, they estimated that a single additional recommendation could improve the sales rank of a book by 3\%. Although plausible in light of the results from experiments, this estimate is likely confounded by other sources of unobservable demand, hence it does not rule out that users would have arrived at the recommended books by some other means in the absence of recommendations.  Oestreicher and Sundararajan~\citeyearpar{oestreicher:2012} and Lin et al.~\citeyearpar{lin:2013} attempted to deal with this problem in a similar manner, studying books on Amazon and digital camera equipment on a Chinese e-commerce site respectively, by constructing sets of ``complementary'' products that were not recommended from the focal product but were likely to experience similar (unobserved) demand.  
Finally, Carmi et al.~\citeyearpar{carmi:2012} and Kummer~\citeyearpar{kummer:2013} also use sets of complementary products to establish conditional independence of demand to the focal and recommended products, but instead exploit exogenous shocks to identify casual effects of recommendations: Carmi et al.~\citeyearpar{carmi:2012} treat appearances on Oprah and in the New York Times Book Review as shocks to demand for books on Amazon, while Kummer~\citeyearpar{kummer:2013} treats natural disasters and listings on the front page of Wikipedia as shocks to the corresponding Wikipedia pages.

In general, the non-experimental papers find large effects of recommendations; for example, 
Oestreicher and Sundararajan estimated that a recommendation amplified demand covariance between otherwise complementary books as much as three-fold. Although this effect seems large relative to the results from experiments, it is hard to compare with them in part because it is expressed in terms of covariance of demand instead of actual demand, and in part because the demand itself is estimated from sales rank using a model~\cite{chevalier:2003}. More importantly, the assumption that the complementary sets do indeed experience the same demand as the recommended sets is critical to their results but ultimately difficult to verify.

Our contribution clearly belongs to the non-experimental category; however, it differs from previous work in three important respects. First, in contrast with rank-based proxies for overall demand used in many of the above studies,
pageview volume from browser logs provides a direct and easily interpretable measure of demand. Second, in contrast with identification strategies that attempt to establish independence of demand for focal and recommended products indirectly, either by conditioning on observable covariates or by comparing correlations with complementary products, our strategy simply controls for demand on recommended products by selecting shocks for which direct traffic to recommended products is known to be constant (and therefore uncorrelated with the focal product). Finally, whereas previous work selects exogenous shocks by first imagining plausible scenarios (e.g., an appearance on Oprah driving traffic to Amazon, or a natural disaster driving traffic to Wikipedia) and then checking for impact, we can measure impact directly from browsing logs, thereby increasing the number and diversity of natural experiments to be analyzed.

\section{Data}
\label{sec:data}

The log data we examine comes from Internet Explorer users who have installed the Bing Toolbar and have explicitly agreed to share their browsing history through it. For each such user, the Bing Toolbar records every URL loaded by the user's browser along with a timestamp and an anonymized user identifier (no personally identifying data is stored in the user logs).
Thus if a user with Bing Toolbar installed visits a product page on Amazon.com, the associated URL will be recorded in the logs. Moreover, because each Amazon URL contains a referral code that identifies the type of link by which the user arrived at the focal page, we can identify whether a user came to a given product through Amazon's search service, a recommendation from another product page, through other Amazon pages (such as a user's cart or wishlist), or via an external website. We can also use these referral codes to infer the (active) network of recommended products on Amazon from browsing logs.
In this manner, we can reconstruct all product pageviews along with the corresponding click-throughs for all Amazon.com user sessions initiated by Bing Toolbar users\footnote{Pageviews that are encrypted via https are logged but not identified, hence we cannot reliably identify purchases, changes to account details, or other secure transactions.}.

\begin{figure*}[t!]
\begin{center}
\includegraphics[width=\textwidth]{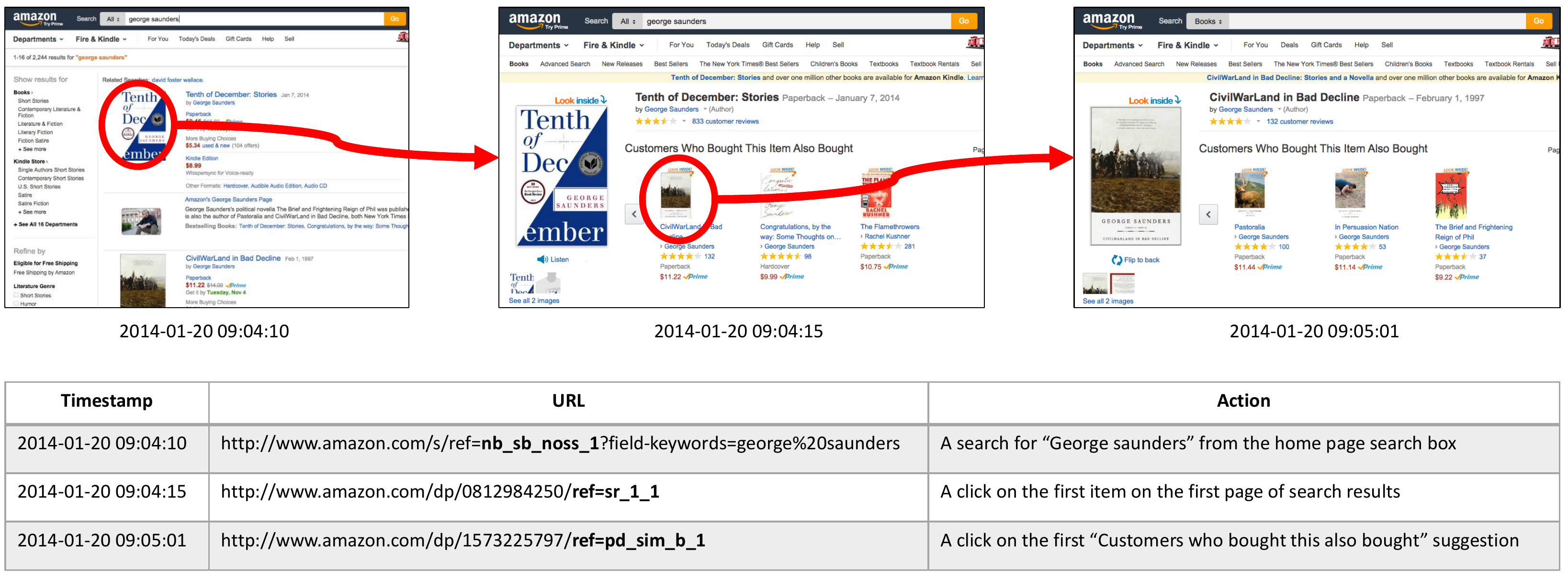}
\caption{Screenshots of an example session and the corresponding logs.}
\label{fig:amazon_screenshot}
\end{center}
\end{figure*}

To illustrate, consider the user session depicted in \fref{fig:amazon_screenshot}.
The first URL we see indicates a search for ``George Saunders''.
The referral code \url{ref=nb_sb_noss_1} contained in this URL specifies that the user issued this search from the home page.
The next URL is for ``Tenth of December'', one of Saunders' books.
Here the referral code \url{ref=sr_1_1} indicates that the user clicked on the first item on the first page search results.
The final URL we see is for ``CivilWarLand in Bad Decline'', another of Saunders' books.
Its referral code, \url{ref=pd_sim_b_1}, indicates that the click came from the first item on the ``Customers who bought this also bought'' list of the previous page.
The presence of referral codes allows us to separate product traffic into two distinct channels: ``direct'' views, defined as traffic that comes from direct browsing or search, such as the first and second pageviews in the example above; and ``recommendation'' views that come from clicks on recommended items, such as the third pageview in this example. The latter also indicates links between a product and its recommendations.
This distinction between direct and recommended visits is critical to our strategy for identifying natural experiments, described in ~\sref{sec:methods}, and hence for obtaining causal estimates about the impact of recommendation systems.

We compiled Amazon session data over a nine month period from September 1, 2013 to May 31, 2014, where to ensure reliable product data, we considered only products that received at least 5 visits over the study period and that were accessible through Amazon's product API. 
We also limited our attention to actual consumer activity by pruning out visits by bots, sellers, or merchants on the Amazon platform.
To eliminate bots, we first removed users who had upwards of 100 visits per day over the entire nine month period.
Next we filtered out users with more than five visits to the \url{sellercentral.amazon.com} or \url{catalog-retail.amazon.com} subdomains, as they are likely to be Amazon sellers.
Finally, we removed users who visited \url{authorcentral.amazon.com} and \url{kdp.amazon.com}, Amazon's portals for authors and publishers.

In addition to user data, we also collected information about the products from the product API, including each product's current price and category.
Amazon categorizes products using two distinct systems: a general ``Product Group'' and a more specific ``Product Type Name''.
Products are often mis-categorized or have missing information, thus we also restricted our attention to products that belonged to groups and types containing at least 100 distinct items.
In practice, this restriction eliminated only 4,000 of the 1.38 million considered products---typically misspelled or unusual product categories and Amazon's own line of products, which, incidentally, do not contain any recommendations ---resulting in items for 60 different product categories.

\begin{figure}[t]
\center
\subfloat[]{{\includegraphics[scale=0.45]{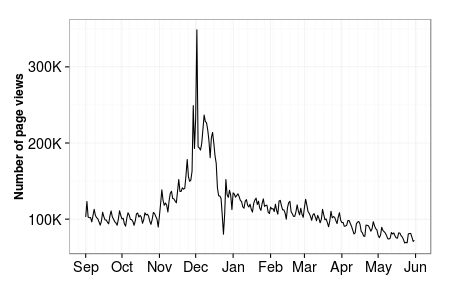} }}%
\subfloat[] {{\includegraphics[scale=0.45]{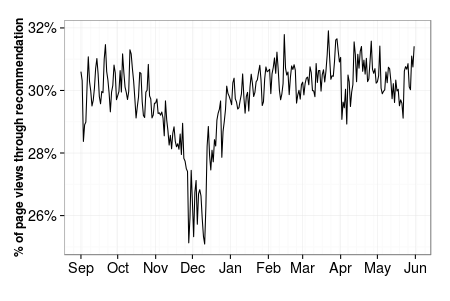}}}%
\caption{Page visits for products on Amazon from September 1 2013 to May 31 2014, both overall (left) and the fraction coming from recommendation click-throughs (right). Note that during the winter holiday season, visits to Amazon go up but the fraction of pageviews from recommendations goes down.}
\label{visits-timeline}
\end{figure}

\begin{figure}[t]
\center
\includegraphics[scale=0.4]{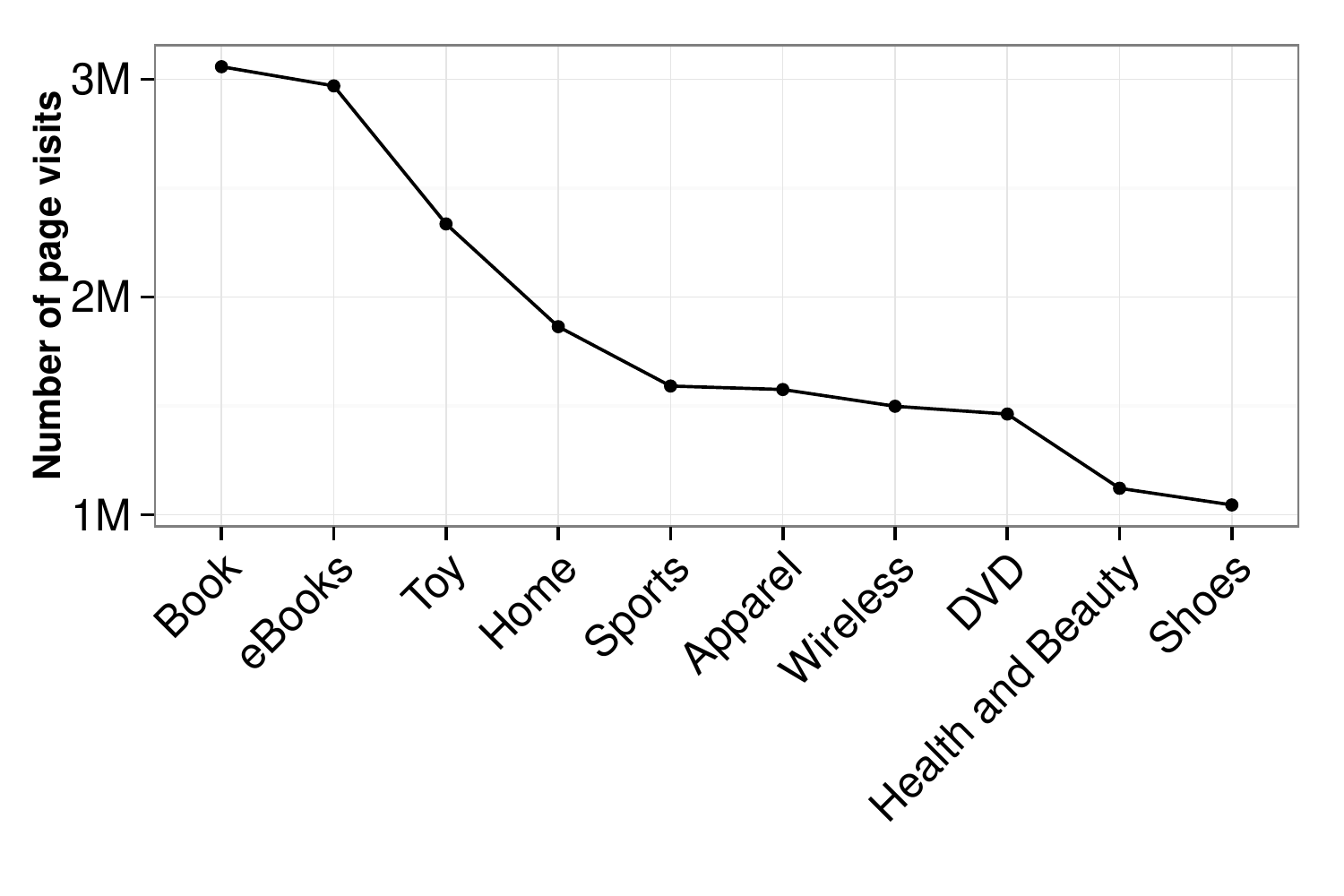}
\includegraphics[scale=0.4]{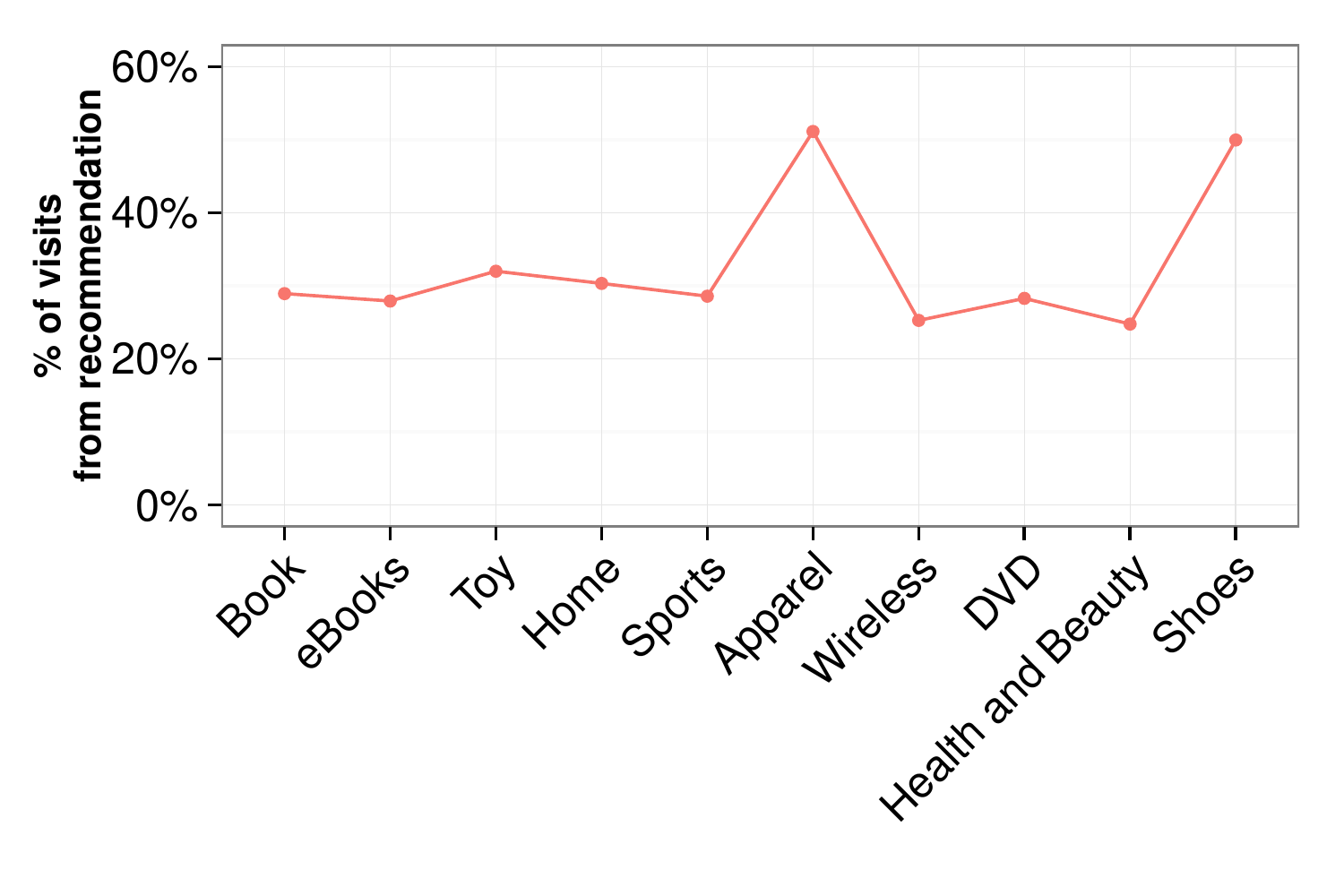}
\vspace{-1.1em}
\caption{Page visits broken down by the product categories of each product. The right panel shows the fraction of page visits from recommendation links. Apparel and Shoes have a higher proportion of visits from recommendations; overall percentage of page visits from recommendations is ~30\%.}
\label{visits-timeline-faceted}
\end{figure}

After filtering on users and items, we are left with 23.4 million visits by 2.1 million users to 1.38 million unique products over the nine-month period of the study. \fref{visits-timeline} and \fref{visits-timeline-faceted} present some basic descriptive statistics of our data, broken down by time and product category respectively.
\fref{visits-timeline} (left panel) shows a timeseries of the total visits to these product pages.
Note that traffic to Amazon peaks during the winter holiday season, in particular on Black Friday and Cyber Monday.
We also observe strong weekly trends, with traffic peaking on Sundays and reaching its lowest on Saturdays.
\fref{visits-timeline} (right panel) shows the fraction of these pageviews that derive from recommendations; i.e., the estimate corresponding to the naive method of counting overall click-throughs. Consistent with previous such estimates~\cite{mulpuru:2006,grau:2009,sharma:2013}, we see an overall trend of roughly 30\% of traffic through recommendations, dipping to about 25\% during the holiday season.
A possible explanation for this dip is that holiday shoppers are looking for specific gifts and are marginally less interested in browsing to discover new items.

\fref{visits-timeline-faceted} (left panel) shows total pageviews for the ten most popular product categories. We see that books and ebooks account for a substantial fraction of traffic, whereas apparel, DVDs, and shoes are less popular but still receive over a million views in the time period.
The right panel shows the percent of traffic derived from recommendations, again broken down by product categories.
We see that total traffic from recommendations varies from just under 30\% of traffic for books and ebooks to over half of all traffic for shoes and apparel.
These differences across categories might reflect that users are discovering more products in these categories, or simply that recommended products have correlated demand.

Before proceeding, we note that the right-hand panels of \fref{visits-timeline} and \fref{visits-timeline-faceted} count the fraction of {\it incoming} traffic to product pages referred through {\it all} types of recommendations on Amazon.
Although this is a natural quantity to measure for a naive estimate of the recommender's impact, in the remainder of the paper we focus our attention instead on {\it outbound} click-throughs from product pages.
The reason is that our identification strategy, described below, relies on estimating the outbound click-through rate on products that receive sudden shocks in traffic.
In addition, we also limit our attention to Amazon's ``Customers who bought this also bought'' recommendations, corresponding to the \url{ref=pd_sim} referrer code, as in \fref{fig:amazon_screenshot}.
These recommendations not only capture the majority of outbound product page clicks, but are also consistently defined across product categories and easily normalized by the number of pageviews to corresponding products. 
Although it makes sense for our method to focus on outgoing traffic for a single type of recommendation rather than incoming clicks of all types, we note that  the naive estimates that we report in \sref{sec:results} will be somewhat lower than in \fref{visits-timeline} and \fref{visits-timeline-faceted}.


\section{Methods}
\label{sec:methods}
In this section we derive our formal identification strategy, specifying the assumptions and conditions under which we can estimate causal click-through rates from observational data.
Specifically, in ~\sref{sec:identification} we present a simple structural model that decomposes recommendation clicks into \emph{causal clicks} and \emph{convenience clicks}, demonstrating the general difficulty in obtaining causal estimates.
We then show that causal estimates are possible when products receive shocks but their recommendations do not.
Finally, in ~\sref{sec:shocks} we describe a set of heuristics to identify such shocks from logged data, and apply these heuristics to browsing data on Amazon to obtain over 4,000 such shocks.

\subsection{Identification Strategy}
\label{sec:identification}
We would like to estimate the impact of a recommender system as measured by the number of additional product pageviews it generates compared to a hypothetical state of the world in which the recommender does not exist.
Estimating this impact from purely observational data is non-trivial because, although we expect that traffic would decrease without the recommender, we do not know the extent to which users might find products through other channels (e.g., search).
Furthermore, it is challenging to separate the effects of a recommender's impact from the inherent demand for recommended products.
As we show below, one strategy for dealing with these difficulties is to look at products that experience instantaneous shocks in traffic while the products recommended next to them do not, thus controlling for confounding factors that might drive interest (and therefore traffic) to both products irrespective of the recommender.

In the language of the causal inference literature~\cite{angrist:2008,dunning:2012,morgan:2007}, our approach is equivalent to an instrumental variable estimate of the click-through rate, where the shock is the instrument, the treatment is exposure to the focal product, and the outcome is click-through to the recommended product. As is typical for instrumental variable approaches, moreover, our causal estimate is not the \emph{average treatment effect} (ATE) that one would obtain from an ideal randomized experiment, but rather a \emph{local average treatment effect} (LATE) that, strictly speaking, estimates the effect only for users who respond to shocks, which in turn is unlikely to be a random sample of the overall population. As~\cite{imbens:2009} has argued, however, the ``LATE vs. ATE'' issue is unavoidable for instrumental variable approaches; thus, in the absence of a randomized experiment on the Amazon website a local, shock-based strategy such as ours is still useful for identifying causal effects provided that the associated concerns regarding generalizability are adequately addressed.


To formalize this idea, consider a focal product, indexed by $i$, and a recommended product shown along side of it, indexed by $j$\footnote{Although we consider specifically ``Customers who bought this also bought" recommendations, the method presented is independent of the underlying recommender algorithm, as long as we obtain click-throughs from a focal product $i$ to its recommended products.}.
Each product has some unobservable demand, specific to that product and possibly varying over time, which we denote by $\uit$ and $\ujt$, respectively.

Although we cannot observe demand directly, we can observe close proxies for demand---namely total views of the focal product, $\vit$, and for the recommended product, $\vjt$. Views of the recommended product can be further broken down into direct visits (e.g., through search or browsing), $\djt$, recommendation click-throughs from the focal product, $\rijt$, and click-throughs from other products that recommend product $j$:
\begin{equation}
  \vjt = \djt + \rijt + \sum_{k \neq i} \rkjt,
\end{equation}
where by restricting our attention to only products $i$ and $j$, we can ignore the third term.
Our identification strategy then hinges on the idea that observing a large change in $\vit$ while $\djt$ remains constant enables us to count views of $j$ that would not have occurred in the absence of recommendations by measuring corresponding changes in $\rijt$.

\begin{figure*}[t]
\begin{center}
\includegraphics[width=\textwidth]{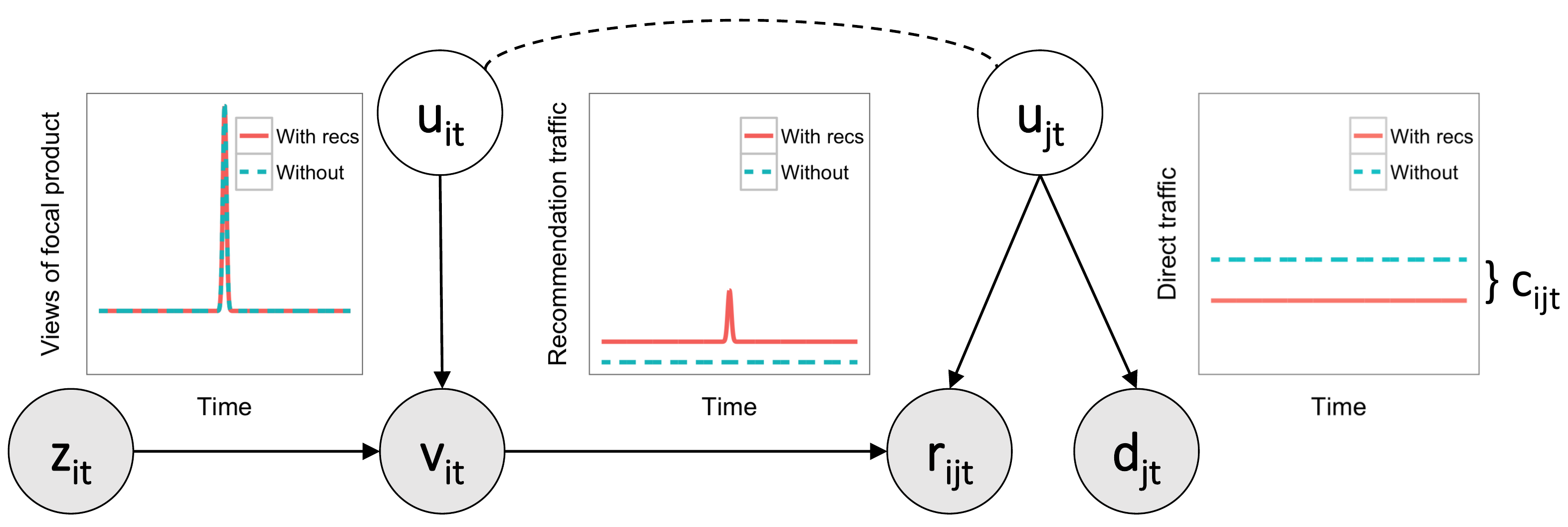}
\caption{A causal graphical model showing the relationships between demand, product views, and recommendation click-throughs.}
\label{fig:causal_diagram}
\end{center}
\end{figure*}

\fref{fig:causal_diagram} depicts the relationships between these variables in a causal graphical model~\cite{pearl:2000} along with illustrative sketches of how they might change over time during a shock, both with and without recommendations present.
The demands $\uit$ and $\ujt$ are unshaded to indicate that they are unobserved, and the dashed line between them indicates that they might, in general, be correlated.
The total traffic to the focal product, $\vit$, is shaded, indicating that it is directly observed, and is composed of observed traffic from an external shock, whose presence/absence is indicated by a binary variable $z_{it}$, as well as from unobserved demand $\uit$. By contrast, the number of direct views to a recommended product, $\djt$, is determined exclusively by $\ujt$ (i.e., the shock applies only to the focal product $i$).
Finally, the number of recommendation click-throughs, $\rijt$, depends on both the traffic to the focal product and the demand for the recommended one.
Thus, when we observe generic changes in $\rijt$ we cannot rule out the possibility that they were driven by a fluctuating  interest in $j$ as opposed to a change in views on $i$---i.e., a ``backdoor pathway''~\cite{pearl:2000} exists from $\vit$ to $\rijt$ via $\uit$ and $\ujt$.

Now consider a hypothetical state of the world in which we remove the recommendation for product $j$ from product $i$'s page.
This change would eliminate recommendation click-throughs $\rijt$, but it might also cause a rise in direct traffic to product $j$, as users who are already aware of or interested in the product make the effort to find it by some other means.
Formally, we define {\it convenience views} $\vpit$ as the number of visits to the focal product $i$ that might have corresponded to views of $j$ in the absence of the recommendation and $\sigmaij$ as the correspondence rate.
The number of {\it convenience clicks} $\cijt \equiv \sigmaij \vpit$ is therefore the potential increase in direct views of product $j$ when the recommendation is removed, as depicted in the far right plot of \fref{fig:causal_diagram} by the difference in the blue and red lines.
Convenience clicks, in other words, represent instances where the recommender merely provides an easier way for users to arrive at a product that they would have otherwise found.
We can now decompose the total number of recommendation click-throughs into causal and convenience clicks as follows:
\begin{equation}
  \rijt = \underbrace{\rhoij \left( \vit - \vpit \right)}_{\mathrm{causal}}~+ \underbrace{\sigmaij \vpit}_{\mathrm{convenience}},
\label{eqn:rijt}
\end{equation}
where $(\vit - \vpit)$ represents views of the focal product $i$ that, by definition, could not have led to views of product $j$ without the recommendation, and $\rhoij$ is the causal click-through rate that we wish to estimate.
From Equation~\ref{eqn:rijt}, it follows that estimating the causal effect of the recommendation system reduces to estimating $\rhoij$; however, we also see that the estimate is confounded by the unknown number of convenience views $\vpit$.

The key to our identification strategy, therefore, is that we limit our attention to recommended products with constant direct traffic $\djt$, so that the demand for these products---and therefore the number of associated convenience views $\vpit$---is known to be constant (see also the red time series sketches in \fref{fig:causal_diagram}).
Moreover, by considering only the \emph{variation} in recommendation click-throughs over time, we eliminate terms proportional to $\vpit$ in \eref{eqn:rijt}, thereby identifying the causal click-through rate:
\begin{eqnarray}
  \rhoij = \DfDx{\rij}{t} / \DfDx{\vi}{t}.
  \label{eqn:rho_hat}
\end{eqnarray}
In the language of instrumental variables, observing constant traffic to the recommended product provides support for the {\it exclusion-restriction} requirement~\cite{dunning:2012}, which states that the instrument (the shock) impacts the outcome (click-throughs) through only the treatment (exposure to the focal product).
Correspondingly, \eref{eqn:rho_hat}, also known as the Wald estimator for a binary instrument~\cite{wald:1940}, estimates the {\it local average treatment effect} (LATE) of the recommender~\cite{imbens:2009}, which in our setting amounts to the causal click-through rate on recommendations for users who participate in shocks.

\subsection{Shocks}
\label{sec:shocks}
In theory we could evaluate \eref{eqn:rho_hat} at any point in time and for any focal product $i$. In practice, however, we limit our attention to focal products that experience a large and sudden shock in traffic at some time $\ts$, for two reasons.
First, as we show later, click-throughs are relatively rare, hence small changes in $\vit$ often do not correspond to observable difference in $\rijt$; thus large changes are necessary in practice to estimate the number of click-throughs.
And second, sudden changes (i.e., shocks) limit potential variability in other elements of the web ecosystem (e.g., a change in search rankings) that might affect the relationship between unobserved demand $\ujt$ for product $j$ and observed traffic $\djt$, and hence might undermine our assumption that constant $\djt$ implies constant $\ujt$.

We operationalize these requirements by looking for days on which a product receives more than 5 times its median daily pageviews over the nine month period.
To ensure that these shocks are sudden, we further require that these high-volume days also show at least a 5-fold increase in traffic over the previous day and at least a 5-fold increase in traffic over the mean daily pageviews over the previous week.
In addition, we require that each shock should contain visits from at least 10 unique users (a filter against events due to a few users' repeated visits), and restrict our attention to products that have at least 5 days of non-zero pageviews within a 14 day window before and after the shock day (to remove ``one-day wonders,'' products without enough data except on shock day).

To summarize, shocks must meet the following criteria, where $\ts$ denotes the time of the shock, $\tb$ indicates the day before the shock, and $\tz$ corresponds to one week earlier.
\begin{itemize}
  \item Visits during the shock must exceed 5 times median traffic: $\vits \ge 5 \cdot \mathrm{median} (\vit)$

  \item Visits during the shock must exceed 5 times the previous day's traffic and 5 times the mean of the last 7 days: $\vits \ge 5 \cdot \vitb$ and $\vits \ge 5 \cdot \mathrm{mean}_{\tz \leq t < \ts} (\vit)$

  \item Visits from at least 10 unique users during the shock

  \item Non-zero visits for at least five out of seven days before and after the shock
\end{itemize}

\begin{figure}[t]
\center
\subfloat[] {{\includegraphics[scale=0.105]{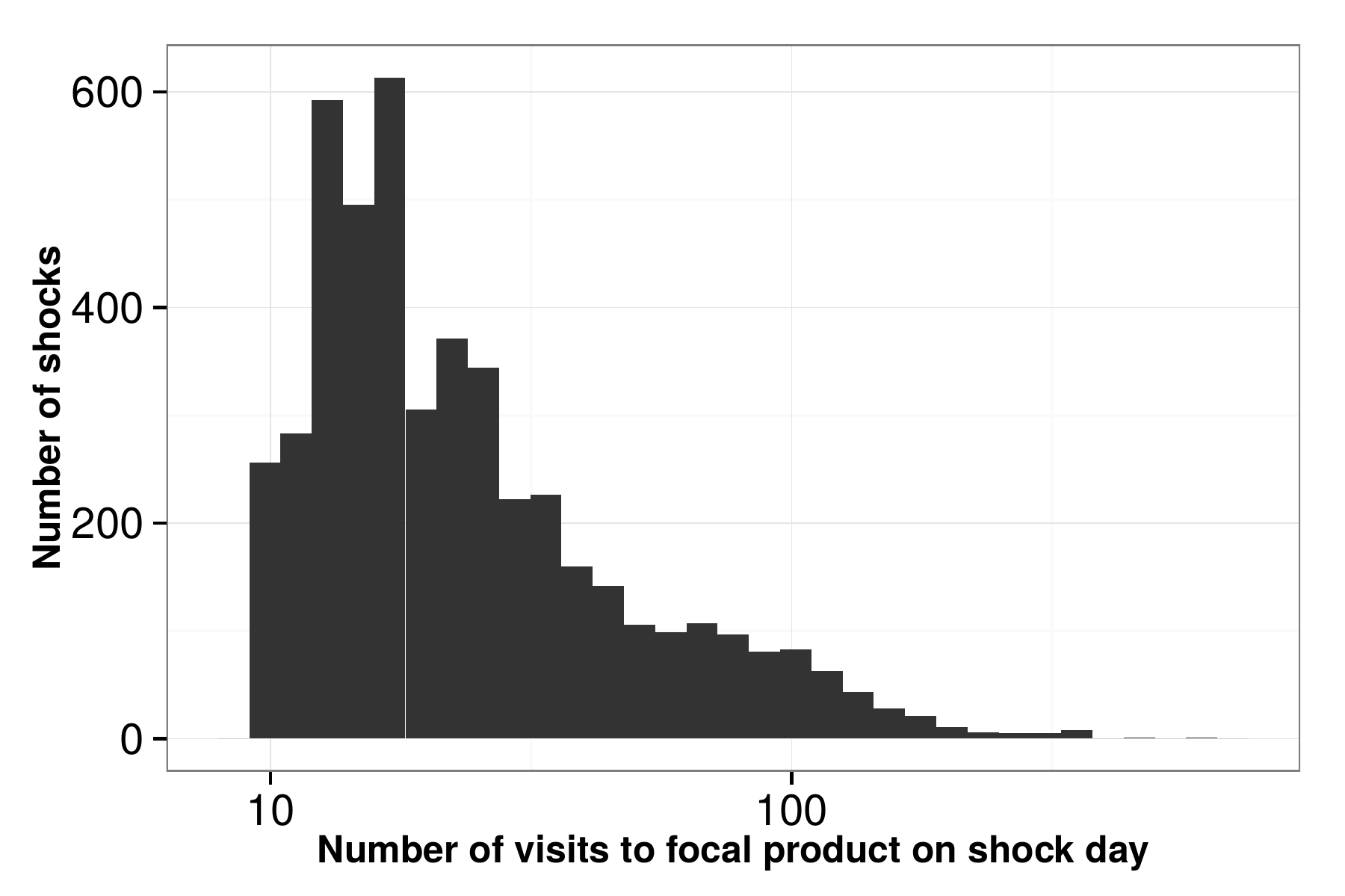} }}%
\subfloat[] {{\includegraphics[scale=0.105]{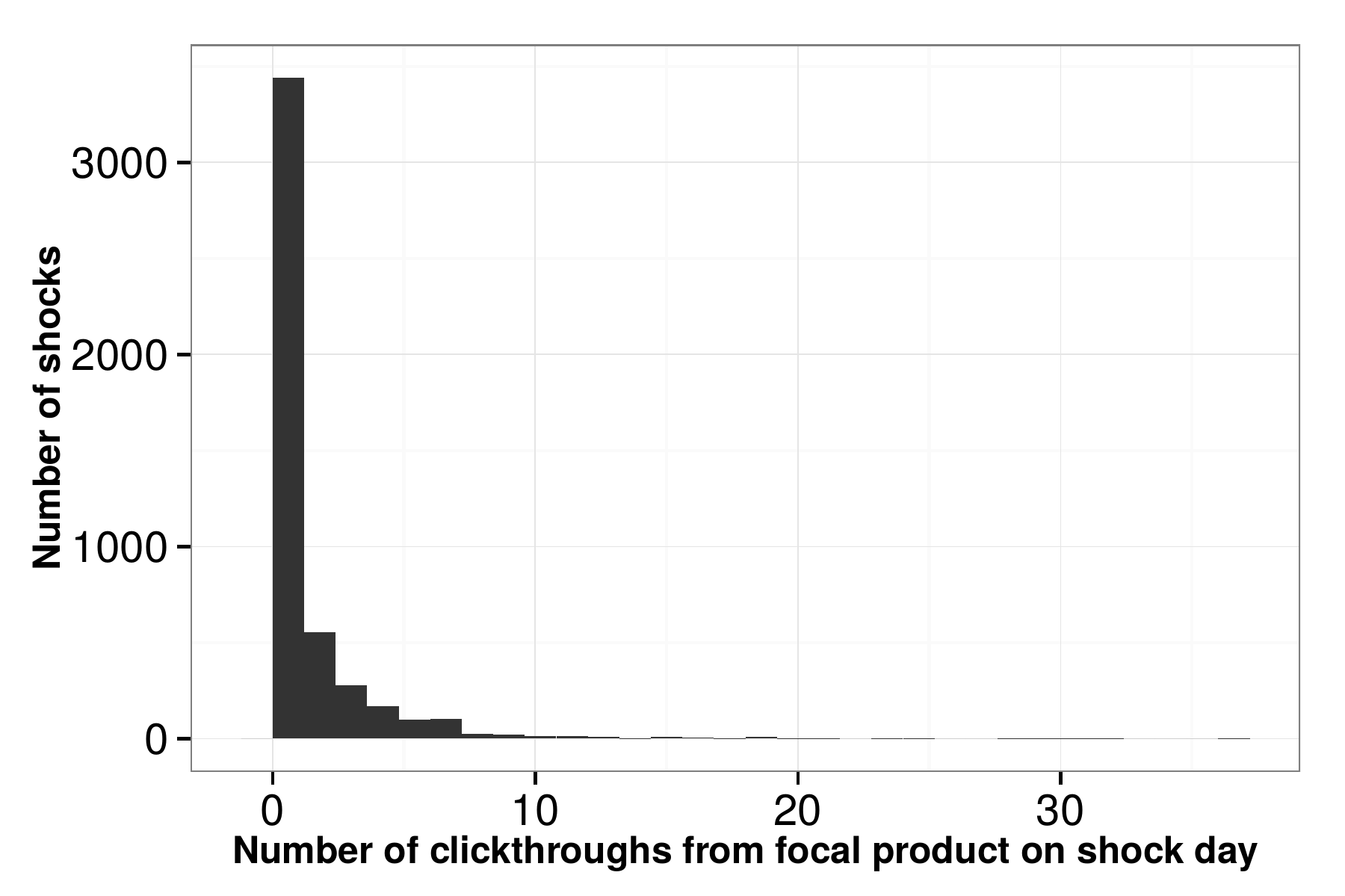} }}
\caption{The distribution of activity on focal products on shock day. The left panel shows the number of page visits to a focal product and right panel shows number of recommendation click-throughs from each focal product. Most shocks do not lead to any click-throughs.}
\label{fig-shock-size-distr}
\end{figure}

When applied to our browsing data, these criteria yielded 4,774 shocks to 4,126 distinct products\footnote{Although these criteria are straightforward and, as we will show later, yield shocks that correspond to our intuition regarding the desired natural experiment they are also clearly arbitrary, at least to some extent. To ensure that our findings are not overly susceptible to the details of our selection criteria, therefore, we also explored a variation in which shocks were required to exhibit 10 times the median traffic and 10 times the previous day traffic. Unsurprisingly we find that these stricter conditions yielded smaller samples of shocks; however, they did not qualitatively alter our results, hence we report only on the more expansive criteria above.} (some products receive mutiple shocks, on different days.)
The left panel of \fref{fig-shock-size-distr} shows the distribution of shock sizes across these products, and reveals that most of the shocks have fewer than 100 visits, with the biggest shock generating 628 visits in one day.
The right panel depicts the distribution of recommendation click-throughs that result from these shocks.
As mentioned above, recommendation click-throughs are relatively rare: even for shocks, we find that a large fraction of focal products have no recommendation click-throughs at all on the day of the shock.

In addition to identifying large and sudden shocks on visits to a focal product, our identification strategy requires that recommended products exhibit stable demand during the time of the shock.
We enforce this condition in practice by requiring that the fluctuation in recent direct visits to recommended products be small in comparison to the size of the shock to the corresponding focal product:

\begin{equation}
  \max_{\tz \leq t \leq \ts} (\djt) - \min_{\tz \leq t \leq \ts} (\djt)
  \le (1-\beta) (\vits - \vitb),
\label{eqn:beta}
\end{equation}
where $\ts$ denotes the time of the shock, $\tb$ indicates the day before the shock, and $\tz$ corresponds to one week earlier.
The parameter $\beta$ allows us to tune the strictness of the constant demand requirement: when $\beta$ is 1, direct visits to recommended products $j$ are exactly constant for the week before the shock, whereas when $\beta$ is 0, variation in direct traffic to recommended products could be as large as the change in traffic due to the shock. 
Theoretically, $\beta = 1$ represents the ideal setting for causal identification. In practice, however, the bulk of shocks that pass this test exhibit so little traffic to $j$, either directly or from click-throughs during the shock or preceding it, that we are unable to estimate $\rhoij$ reliably. At the other extreme, meanwhile, manual inspection of shocks allowed by $\beta = 0$ reveals that they frequently violate any plausible interpretation of constant demand for product $j$, hence fail to satisfy the assumptions of our identification strategy. 

\begin{figure}[t]
\center
\includegraphics[scale=0.4]{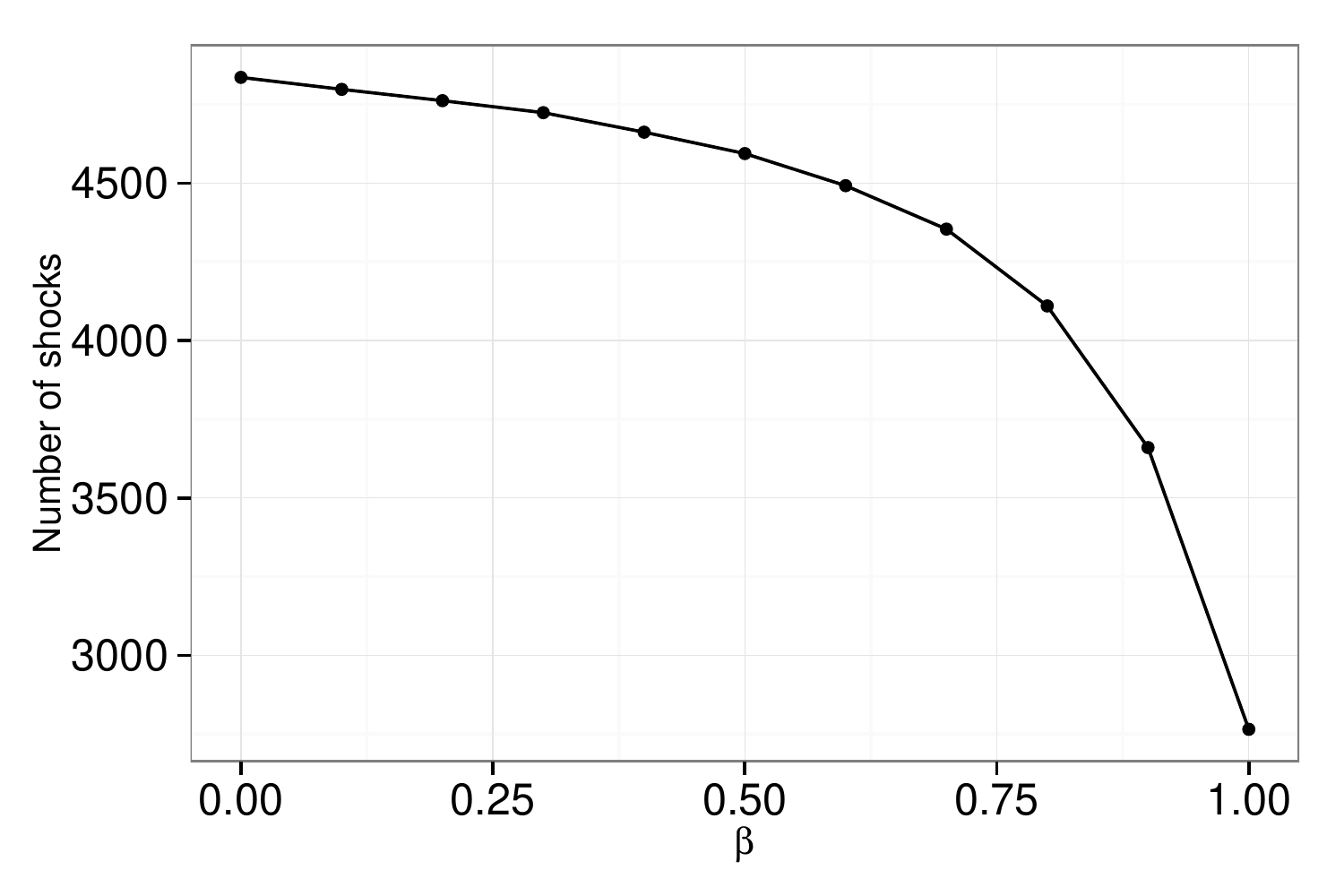}
\vspace{-1.0em}
\caption{The number of remaining shocks as we limit to products whose recommended items have increasingly constant demand.}
\label{fig-beta-num-shocks}
\end{figure}

In practice, therefore, we must choose an intermediate value of $\beta$ that strikes a reasonable tradeoff between \emph{estimation} (determined by volume of traffic) on the one hand and \emph{identification} (determined by constancy of demand for $j$) on the other hand.
\fref{fig-beta-num-shocks} shows the number of shocks remaining as we vary $\beta$ between 0 and 1, where for a given value of $\beta$ we discard any shocked product that has at least one recommended item that violates \eref{eqn:beta}.
As we increase the value of $\beta$, we not only limit our attention to fewer products, but also tend to select for products whose recommendations have lower direct traffic, as per \eref{eqn:beta}.
Using too large a value of $\beta$ would not only leave us with fewer products for reliable estimation, but would also bias our estimated causal rate towards artificially low values.
Noting that 90\% of all shocks (4,314) are retained up to $\beta=0.7$, after which the number of remaining shocks drops rapidly, we concentrate our attention on shocks for which $\beta=0.7$.

\begin{figure}[h]
\center
\subfloat[Accepted shock at $\beta=0.7$]{{\includegraphics[width=0.5\textwidth]{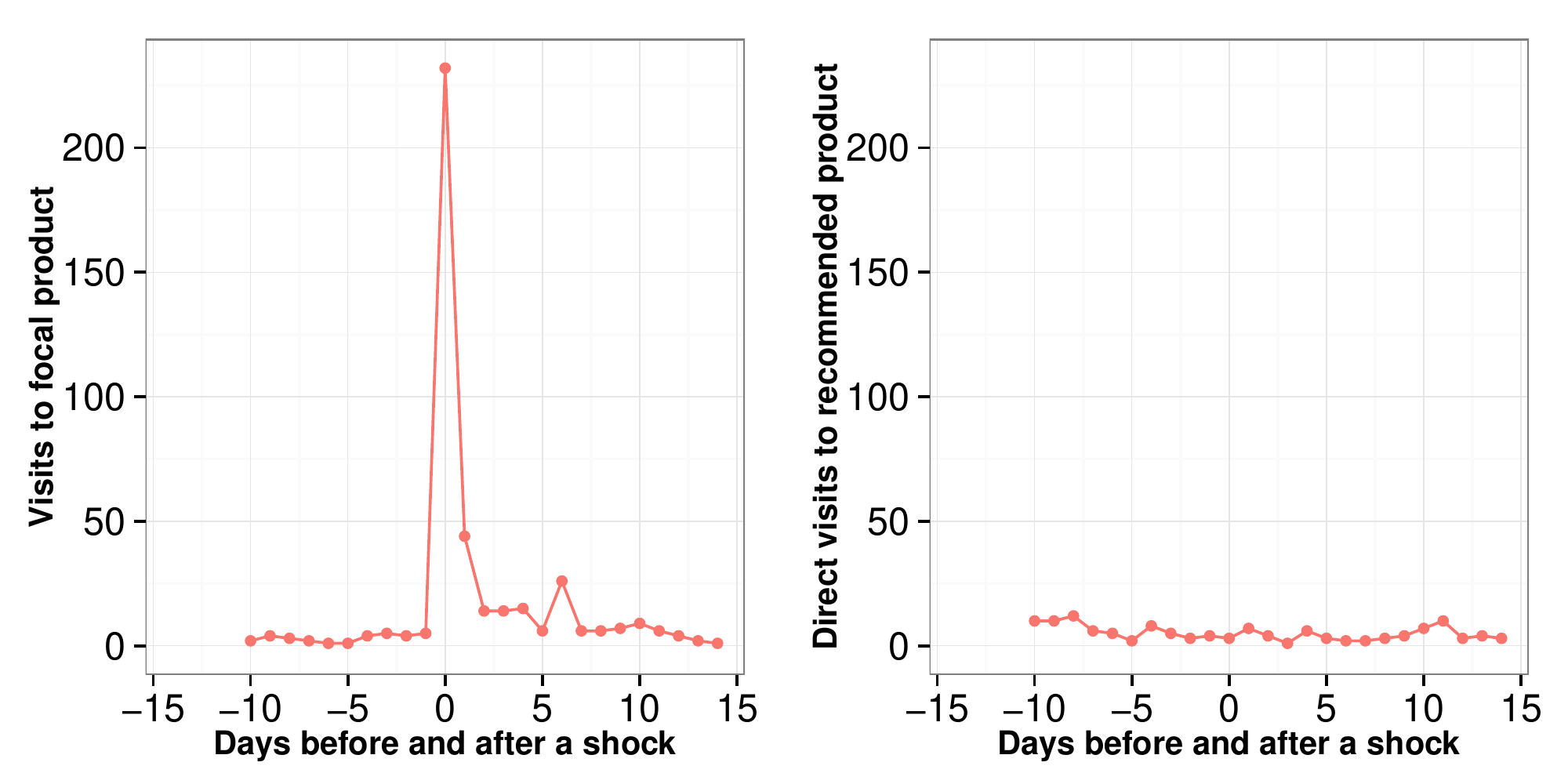} }}%
\subfloat[Rejected shock at $\beta=0.7$] {{\includegraphics[width=0.5\textwidth]{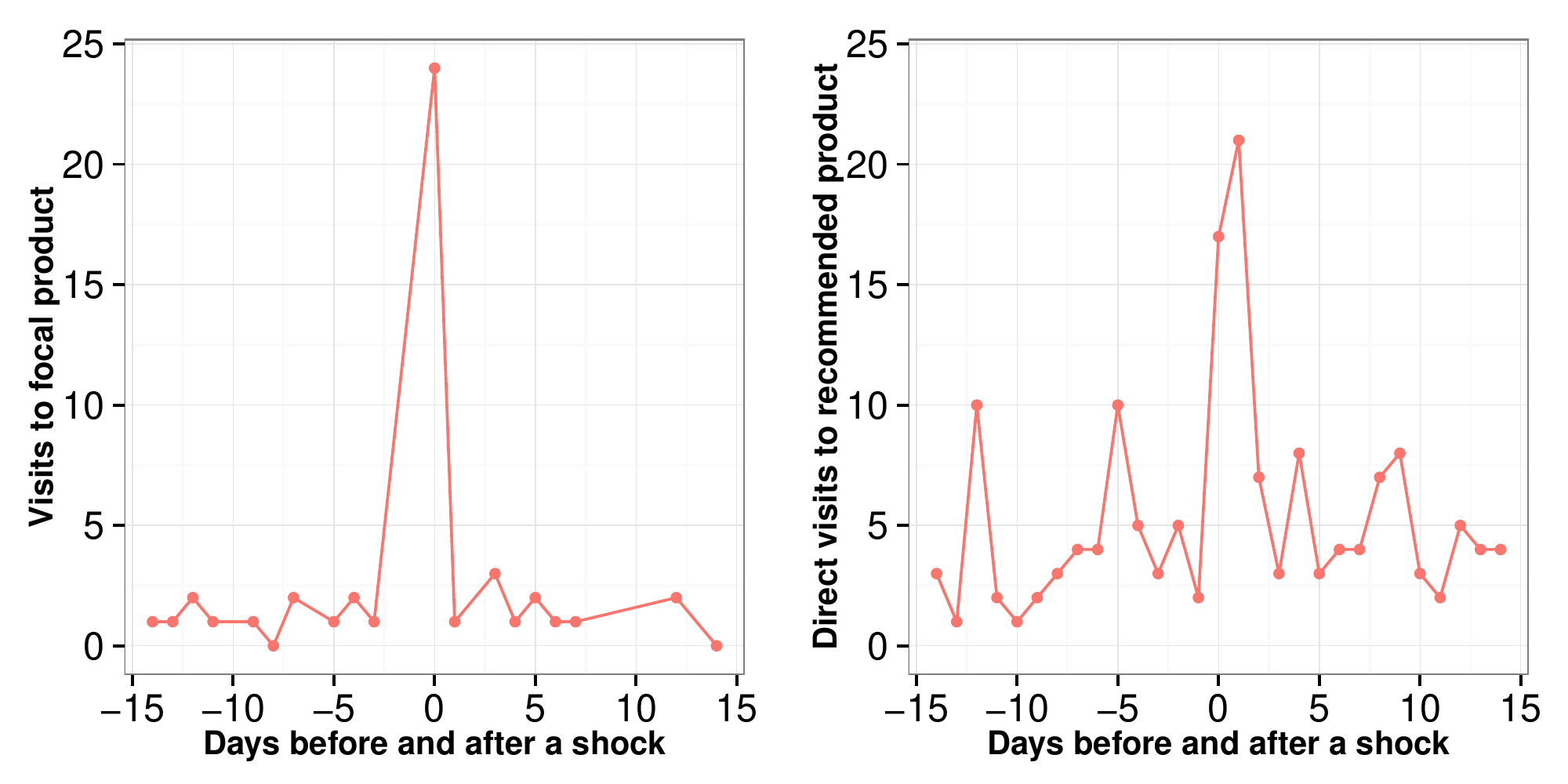}}}%
\caption{Examples of accepted (left) and rejected (right) shocks for $\beta=0.7$.}
\label{fig-shocks-visualize1}
\end{figure}

\fref{fig-shocks-visualize1} illustrates the shock patterns that pass (left) and fail (right) this filter.
In the left panel, the focal product receives an influx of hundreds of visits during the shock, while direct visits to its recommendation vary by a handful of visits.
Although the demand for this recommendation may vary slightly over the time period, it is highly unlikely that this variation is correlated with the sudden increase in interest to the focal product, making causal identification possible.
Contrast this situation to the rejected event in the right panel, where we see that the focal and recommended products receive almost identical increases in traffic at the time of the shock; in other words, a clear example of the type of correlation in demand that hinders causal identification.
From manual inspection, we have verified that these patterns are typical, hence from now on we use the set of shocks corresponding to $\beta=0.7$ as our canonical sample\footnote{As before, we note that our results are quite robust to what method we use to identify shocks and to the choice of $\beta$, so long as $\beta$ is high enough to eliminate correlation and low enough to leave a reasonable number of events for our analysis. For example, we also considered a number of other heuristics for identifying shocks, such as comparing shock to the median (instead of mean) preceding traffic and relaxing the constraint that the shock occur within one day, all with qualitatively similar results.}.

\section{Results}
\label{sec:results}
In this section, we first compute an empirical estimate for the causal click-through rate $\rho$ using the identified shocks in our dataset. From this estimate, we then compute the fraction $\lambda$ of observed recommendation clicks that we estimate to be causal, and further examine this overall rate by product category. Finally, we examine the generalizability of our findings in light of a number of potential sources of non-randomness of our sample of shocked products.

\subsection{Estimating the causal click-through rate}
For each focal product $i$ with an eligible shock, we can now use Equation~\ref{eqn:rho_hat} to compute $\rhoijhat$ empirically as the ratio between change in recommendation clicks on $j$ to change in visits to the focal product due to a shock\footnote{To reduce the impact of duplicate visits to either the focal or recommended product by the same user in the same session, we count a visit to a product in the same session only once. That is, multiple visits to the same focal product are counted as a single visit, and similarly, multiple visits to a recommended product from the same focal product are counted as a single visit.}:
\begin{eqnarray}
	\rhoijhat = \frac{\Delta \rijts}{\Delta \vits},
\end{eqnarray}
where we have approximated derivatives by discrete changes in time. To ensure we have enough data for reliable estimates, we calculate $\rhoihat$, which uses the sum of all outgoing recommendation click-throughs from a focal product:
\begin{eqnarray}
	\rhoihat = \sum_j \rhoijhat.
\end{eqnarray}
Further, we reduce noise in our estimates by considering a window of one week before the shock.\footnote{We choose a one week period, but our results are similar for windows of 3, 5, and 14 days prior to the shock.}

\begin{figure}[t]
\includegraphics[scale=0.4]{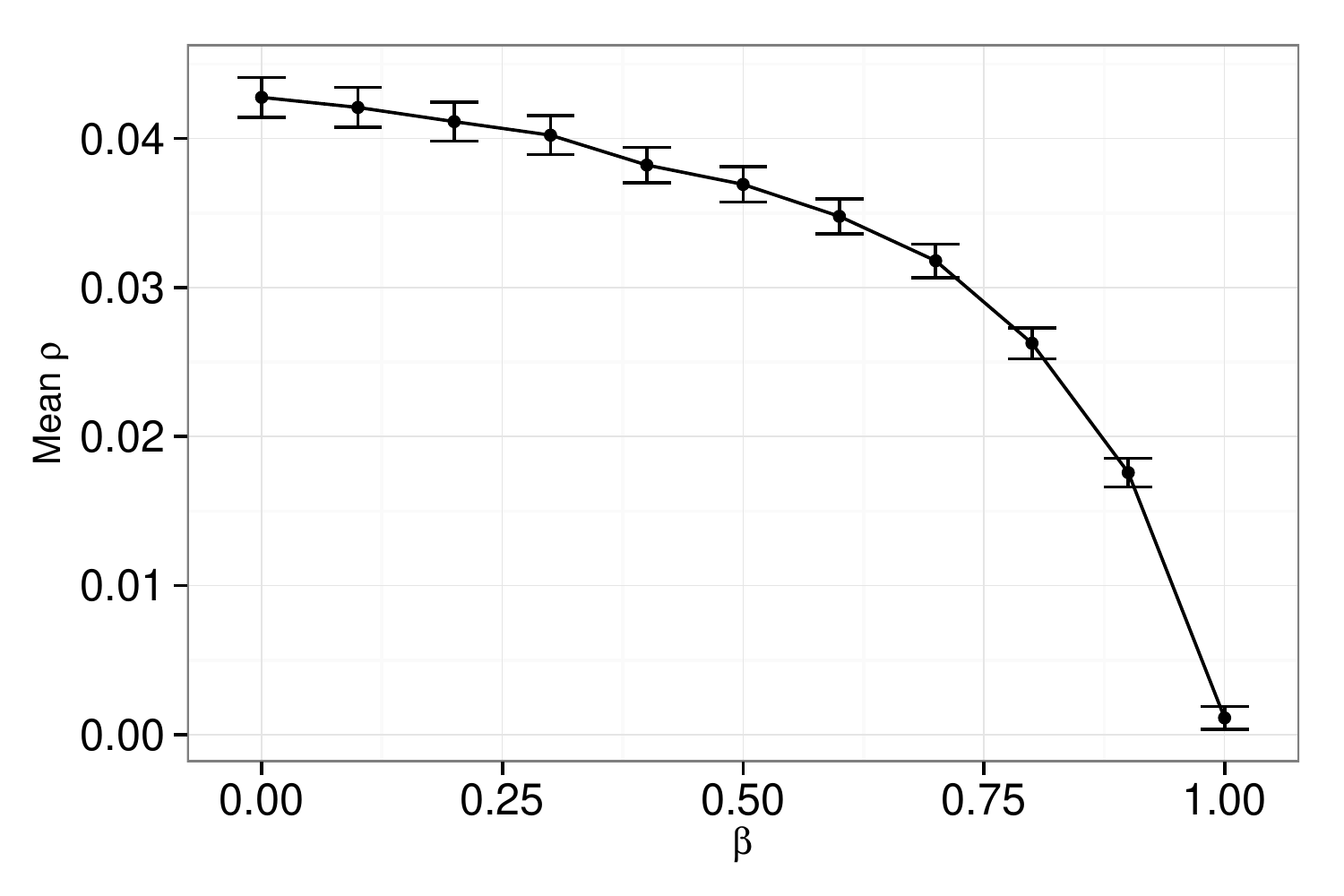}
\includegraphics[scale=0.4]{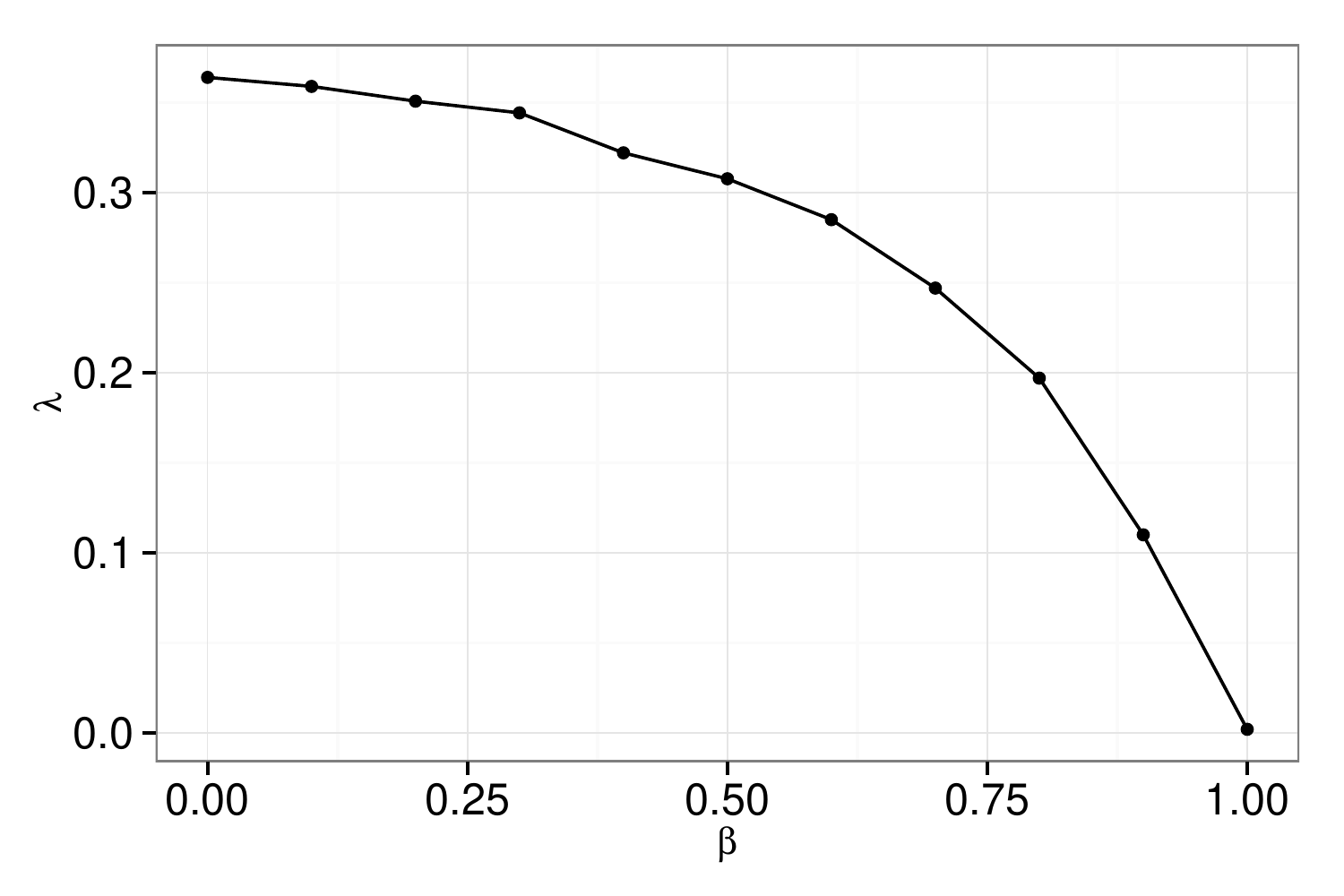}
\vspace{-1.0em}
\caption{The left panel shows the variation in the estimated causal click-through rate as we limit to products whose recommendations have increasingly constant demand. The right panel shows the fraction of recommender traffic estimated to be causal as we limit to products whose recommendations have increasingly constant demand.}
\label{fig:beta-rho}
\label{fig:beta-lambda}
\end{figure}

The left panel of Figure~\ref{fig:beta-rho} shows the estimated click-through rate $\rhoihat$ for different values of $\beta$, where the error bars show standard errors on the estimate. As expected, $\rhoihat$ decreases monotonically with increasing $\beta$ corresponding to an increasingly stringent control on the exogenous demand for the recommended product. This figure suggests that the most generous upper bound on the causal click-through rate is roughly 4\% (for $\beta=0$) while the more realistic estimate, corresponding to $\beta = 0.7$, is closer to 3\%.

\subsection{Estimating the fraction of causal clicks}
We now use this estimated conversion rate to obtain an upper bound on 
the ratio between views caused by the recommender and all observed recommendation clicks prior to the shock:
\begin{equation}
  \rhoihat \vitb \ge \rhoihat \left( \vitb - \cjtb \right) = \text{causal clicks}
\end{equation}
The bound comes from the generous assumption that there are no convenience views before the shock, so that all click-throughs are causal and determined simply by multiplying the conversion rate $\rhoihat$ by the number of views prior to the shock, $\vitb$. Dividing this upper bound by the observed number of click-throughs $\rijtb$ gives us an upper bound on the fraction of recommender traffic that can be considered causal:
\begin{eqnarray}
  \lambdaij \equiv {\rhoij \vitb \over \rijtb}.
\label{eqn:lambda_def}
\end{eqnarray}
Finding $\lambdaij =1$ would therefore imply that under the most generous assumptions a simple observational estimate of the number of recommendation click-throughs from $i$ to $j$ might capture the impact of the recommendation system, whereas any $\lambdaij < 1$ would necessarily imply that not all observed click-throughs should be attributed to the recommendation system.

\begin{figure}[t]
\begin{center}
\includegraphics[width=0.55\textwidth]{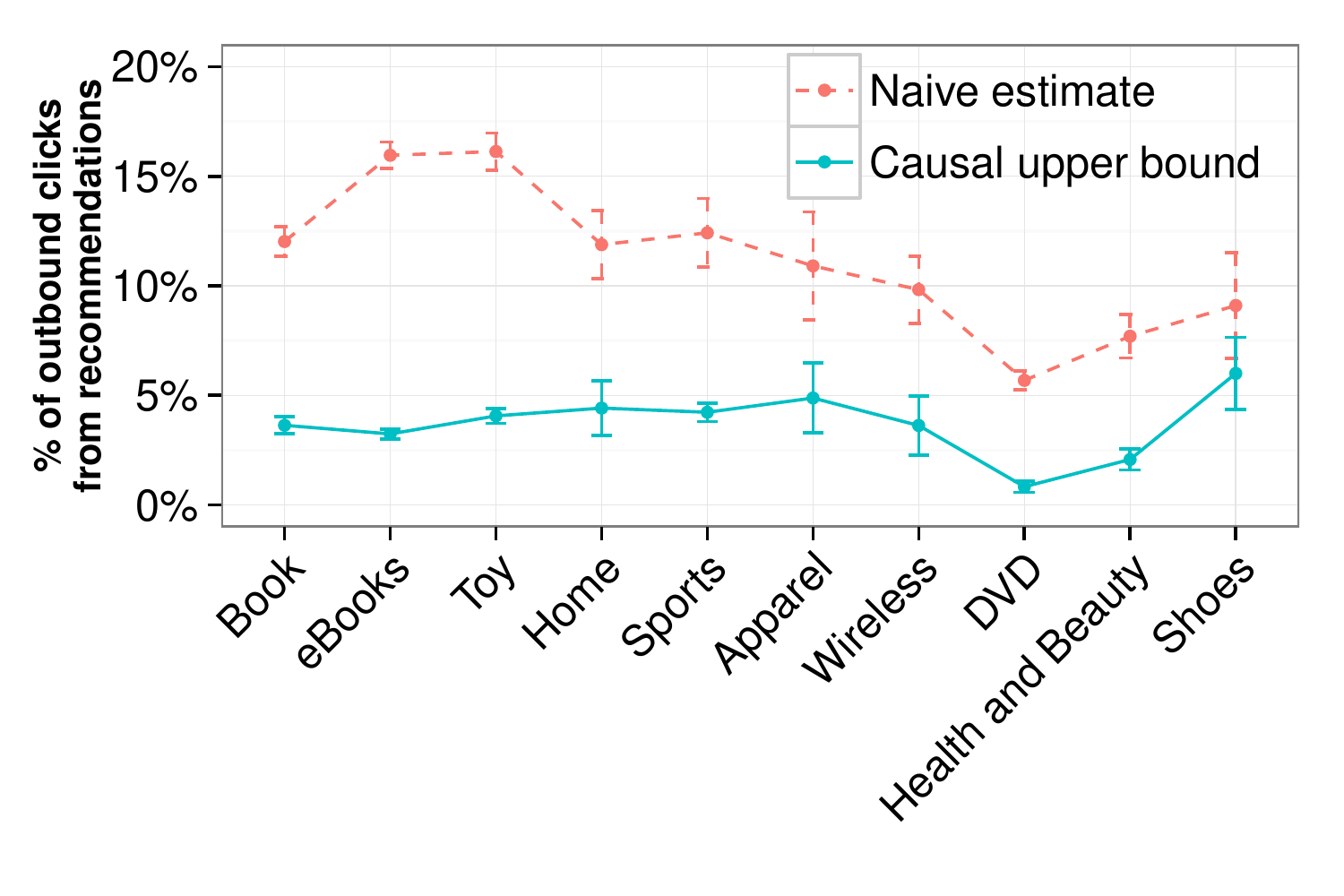}
\end{center}
\vspace{-2.5em}
\caption{Comparison of naive click-through rate for shocked products in the ten most popular product groups on Amazon.}
\label{fig:category-clicks}
\end{figure}

As with $\rhoij$, we reduce noise in our estimate by summing over all focal and recommended products during a time window of one week before the shock to obtain an empirical estimate of the overall fraction of causal clicks, denoted $\lambdaoverall$, as\footnote{We do not compute $\lambda_i$ separately for each shock as most focal products have little traffic to them before the shock, hence $\rijtb$ is frequently zero. Summing first over all focal products eliminates infinite values of $\lambda_i$ while still capturing the overall effect.}:
\begin{eqnarray}
	\lambdaoverall \equiv \frac{\sum_i \sum_{\tz \leq t < \ts} \rhoihat \vit}{\sum_{i,j} \sum_{\tz \leq t < \ts} \rijt}
\end{eqnarray}
The right panel of Figure~\ref{fig:beta-lambda} shows empirical estimates for $\lambdaoverall$ as a function of increasing $\beta$. As with $\rhoihat,$ $\lambdaoverall$ decreases monotonically with $\beta$, where for $\beta = 0.7$ the value is $\lambdaoverall \approx 25\%.$ Overall, therefore, we conclude that only about a quarter of observed traffic from recommendations is actually caused by the recommendation system, while the remainder represents convenience clicks that would have occurred anyway.

Finally, ~\fref{fig:category-clicks} compares the naive estimates of recommendation traffic by product group
with the corresponding causal estimates.
The dashed red line shows the naive estimate, the mean conversion rate on recommendations for shocked products in each category, implying a conversion rate of more than 15\% on recommendations on ebooks and toys, for instance.
The solid blue line, in contrast, shows the average value of $\rhoihat$ for shocked products in each category, indicating that the majority of observed clicks are merely due to convenience, and that a more accurate estimate of the causal impact of the recommender is 5\% or lower across these and other categories.\footnote{Recall that the dashed red line in \fref{fig:category-clicks} differs from the solid red line in \fref{visits-timeline-faceted} because the former looks only at outbound clicks from ``People who bought this also bought'' recommendations, whereas the latter examines inbound clicks from all recommendations.}

\subsection{Generalization issues}
We now address concerns about extrapolating our results to all recommendations shown on Amazon.
Our identification strategy enables us to estimate the causal impact of exposure to recommendations for users who respond to product shocks. As noted earlier, however, our reliance on an instrumental variable approach means that we are estimating a {\it local average treatment effect} (LATE), which may differ from the overall causal impact of the recommender across all users and products, known as the {\it average treatment effect} (ATE).
For instance, products that receive shocks and the users that participate in them may not be a representative sample of traffic on Amazon.
Thus, we examine three major threats to the external validity of our results: price discounts, holiday effects, and distribution of user activity and product popularity for shocks.

\subsubsection{Price discounts on focal products}
Amazon routinely offers deals on its products, thus one of the reasons for shocks on a product could be that it is on sale. Products on sale, moreover, might be expected to be more attractive than usual relative to the recommended products (which would seem relatively more expensive), in which case we would observe an artificially low click-through rate. We would therefore like to reassure ourselves that our sample of shocks is not dominated by flash sales or other price cutting activity.
Unfortunately, it is not possible to get a product's price on a particular date in the past through Amazon's API, so we instead checked for the effect of price indirectly in two ways. 

First, we examined external websites that drive traffic to Amazon and checked for any change in the distribution of their referral share for shocked products. The intuition here is that some of the people who arrive at a discounted product's page would be referred from external channels such as deal websites and e-mail, and thus the sources of traffic during shocks would be different than on normal days.
For all shocked products except for ebooks, we found no significant change in referrers on shock day. For ebooks, traffic from deal websites such as \url{bookbub.com} accounts for approximately 3\% of the total page visits on shocked products, compared to a negligible fraction on other days. We compared our results with and without these visits from deal-specific sites such as \url{bookbub.com}, and found little change. Thus, traffic from deal sites does not appear to alter our findings. Second, we used third party services that keep track of historical prices for popular products on Amazon to look at price variation.
In particular, we used \url{camelcamelcamel.com} to manually inspect the prices of 30 focal products on shock day. (The site does not provide access to an API, and a larger scale analysis would violate terms of service). We did not find a noticeable change in product prices on the day of the shock among the products we examined.

\subsubsection{Holiday effects}
As we saw in ~\fref{visits-timeline}, Amazon receives an increase in overall traffic during the winter holiday season.
To verify that shocks are not all sourced from this period we looked at the temporal distribution of shocks.
Shocks occur throughout the nine month period we studied, but are in fact concentrated in the holiday period.
To test if our results are confounded by a holiday effect, we re-ran our analysis excluding shocks that occurred between November 15th and December 31st, and found no significant differences in results.

\begin{figure}
\begin{center}
\includegraphics[scale=0.45]{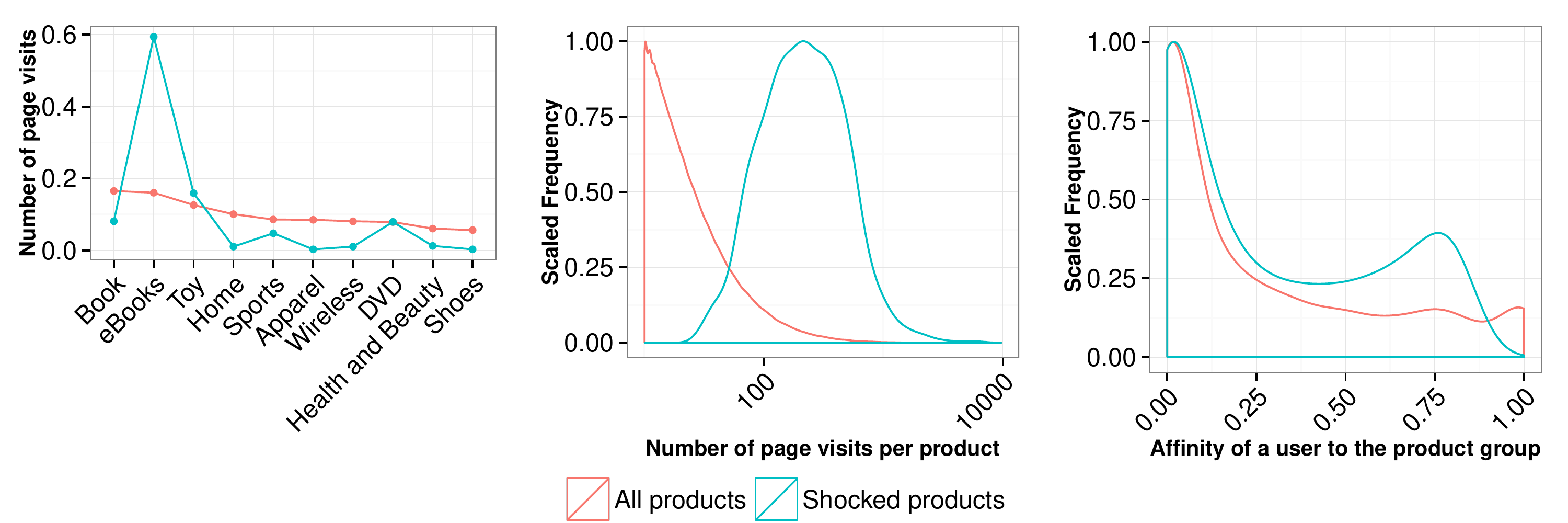}
\end{center}
\vspace{-1.0em}
\caption{Robustness checks.}
\label{fig-robustness}
\end{figure}

\subsubsection{Distribution of user and product attributes}
We also conducted a comparative analysis of shocked products with all other products with respect to product group, product popularity, and user interest.  \fref{fig-robustness} shows the results for these three sets of attributes in turn. First, the left panel of \fref{fig-robustness} compares the distribution of pageviews across product groups for shocked products to the same distribution over all products.
While we find shocks for each of the top 10 product groups, we see that shocks are concentrated among different groups compared to regular activity.
Ebooks, for instance, receive a disproportionate number of shocks relative their usual share of traffic, whereas DVDs are somewhat under-represented in the set of shocks. Although our overall estimate of $\lambda$ is therefore likely biased toward its value for ebooks, ~\fref{fig:category-clicks} indicates that the variation in the causal click-through rate is small among the top-5 product categories, and thus the  potential for error is small.
Second, the middle panel of \fref{fig-robustness} compares the distribution of pageviews to shocked products and all other products, regardless of category.
This highlights that products that receive shocks are, on average, more popular than randomly sampled products from Amazon's catalog. Although this difference is not surprising, given our method for identifying shocks, it nonetheless may introduce some difficulty in generalizing from our sample to the overall population.

A final concern regarding the generalization of activity on shocks could be that users who visit products due to a shock may be very different from the regular visitors to Amazon (thereby also violating as-if random assignment).
Specifically, if users visiting shocked products are unusually (dis)interested in the product compared to routine users, then our estimates might be biased.
To test this, we first computed a preference profile for each user based on the distribution of their pageviews over product groups.
We define a user's ``affinity'' for a product as the fraction of pageviews the user distributes to the product's category, and compute the affinity between each user and each product they visited.
The right panel of Figure~\ref{fig-robustness} shows the affinity distribution for shocked and non-shocked products, from which we see that most visits are to low-affinity products, but shocks have a higher proportion of high affinity visitors compared to typical traffic.
This indicates that users with highly targeted interests are somewhat over-represented during shocks.
All three of these concerns emphasize caution in extrapolating the above results to general activity on all of Amazon, which we discuss below.


\section{Discussion}
\label{sec:discussion}
In this paper, we have presented a method for estimating the causal impact of recommendations using natural experiments.
Our method is both conceptually simple and also practical, requiring only access to separate counts of recommendation-driven traffic and total visits to individual pages over time---data that are readily available to practitioners running their own recommendation systems and inexpensive compared to running A/B tests.
Furthermore, this method can be used to estimate causal click-through rates in more than just recommendation systems---for instance, it could be used to assess the effectiveness of contextual advertisements.
By controlling for direct traffic as a proxy for product demand, our method eliminates the need to fit statistical models or construct comparable product sets to control for unknown product demand, as was necessary in previous work~\cite{garfinkel:2006, oestreicher:2012, carmi:2012, kummer:2013}.
Applying our method to a large-scale dataset of browsing activity on Amazon, we found that only a quarter of recommendation click-throughs on shocked products can be considered causal.

As mentioned above, however, some caution should be taken when extrapolating this result to overall traffic on Amazon, let alone to other websites.
First, we limited our analysis to click-throughs from Amazon's ``Customers who bought this also bought'' recommendations, which are just one of many ways in which the site surfaces recommendations.
In particular, these recommendations are specific to the product page on which they are shown, but are not personalized to the user viewing the page.
Our method can be applied to personalized recommendations as well, but inferred causal rates may differ from those found here.
Second, the shocked products we studied were not a random sample of all products on the site.
Shocked products tended to be relatively popular ones and certain categories (e.g., ebooks) were over-represented among them. Moreover, a fraction of the users who visited shocked products also had unusually high interest in them compared to routine visitors.
Third, as noted in Section ~\ref{sec:methods}, raising the value of $\beta$ (to increase the degree of demand constancy required for recommended products) also effectively restricted our sample to products with less interesting recommendations, driving down the inferred click-through rate.
In principle one can deal with issues of representativeness (which can also arise when trying to generalize the results of randomized experiments) via post-stratification to reweight our causal click-through estimates so that they mirror those of a randomly drawn sample of product visits~\cite{little:1993}.
In practice, however, this would require a larger dataset, as we would have to make (or model) separate estimates across product group, product traffic, and user interest; thus we leave this exercise for future work. 

Finally, we emphasize that the natural experiments we considered estimated the causal impact of recommendations for only one focal product at a time, as opposed to the effect of turning on or off the recommender across an entire site.
As a result, we expect that causal click-through rates produced by our method provide an {\it overestimate} of the overall impact of recommender systems.
If {\it all} recommendations were removed from the entire site, we expect that users would naturally shift to exerting more effort when searching for products, as they would not expect to see related results on product pages.
Although users would certainly discover fewer products if the site did not show any recommendations, they would probably still manage to find the products they are already aware of; thus we regard our estimate as an upper bound on the total causal effect of the recommender.
Given all these caveats, it is nonetheless encouraging that if we apply our estimate of $\lambdaoverall = 0.25$ to all recommendation traffic as described in \fref{visits-timeline-faceted}, we compute that the total fraction of traffic caused by the recommender is roughly 8\%, a figure that is surprisingly consistent with results from a recent randomized experiment~\cite{belluf:2012}.



\vspace{-0.54em}
\begin{acks}
We thank Giorgos Zervas for his assistance in parsing the Bing Toolbar logs, Justin Rao for relevant references and useful discussions on threats to our identification strategy, and Dan Cosley for his feedback on this work.
\end{acks}

\vspace{-0.5em}
\bibliographystyle{acmsmall}




\end{document}